\documentclass[amsmath,amssymb,aps,pra,superscriptaddress,twocolumn
]{revtex4-2}

\usepackage{amssymb}
\usepackage{graphicx}
\usepackage{dcolumn}
\usepackage{bm}
\usepackage{amsmath}
\usepackage[colorlinks,urlcolor=cyan,citecolor=blue,linkcolor=magenta]{hyperref}

\begin{document}

\title{Collisional dynamics of symmetric two-dimensional quantum droplets}
\author{Yanming Hu}
\affiliation{Institute of Theoretical Physics and State Key Laboratory of Quantum Optics and Quantum Optics Devices, Shanxi University, Taiyuan 030006, China}

\author{Yifan Fei}
\affiliation{Department of Physics and Key Laboratory of Optical Field Manipulation of Zhejiang Province, Zhejiang Sci-Tech University, Hangzhou 310018, China}

\author{Xiao-Long Chen}
\email{xiaolongchen@swin.edu.au}
\affiliation{Department of Physics and Key Laboratory of Optical Field Manipulation of Zhejiang Province, Zhejiang Sci-Tech University, Hangzhou 310018, China}

\author{Yunbo Zhang}
\email{ybzhang@zstu.edu.cn}
\affiliation{Department of Physics and Key Laboratory of Optical Field Manipulation of Zhejiang Province, Zhejiang Sci-Tech University, Hangzhou 310018, China}

\begin{abstract}
The collisional dynamics of two symmetric droplets with equal intraspecies scattering lengths and particle number density for each component is studied by solving the corresponding extended Gross-Pitaevskii equation in two dimensions by including a logarithmic correction term in the usual contact interaction. We find the merging droplet after collision experiences a quadrupole oscillation in its shape and the oscillation period is found to be independent of the incidental momentum for small droplets. With increasing collision momentum the colliding droplets may separate into two, or even more, and finally into small pieces of droplets. For these dynamical phases, we manage to present boundaries determined by the remnant particle number in the central area and the damped oscillation of the quadrupole mode. A stability peak for the existence of droplets emerges at the critical particle number $N_c \simeq 48$ for the quasi-Gaussian and flat-top shapes of the droplets.
\end{abstract}

\maketitle

\section{Introduction} \label{Introduction}
The self-binding property of classical liquids has been explained by van der Waals theory: the liquid can exist stably because the long-range attraction and short-range repulsion reach balance under a certain density \cite{margenau1939van}. The Bose-Einstein condensates (BEC) are generally described by the Gross-Pitaevskii equation (GPE) in the mean field approximation and in the self-bound state of a many-body system the mean field term in the two-component condensates provides such an attractive force. A repulsive potential with a power-law dependence on density higher than that of the mean field is required to balance the attraction. The earlier attempt to reach this balance was to utilize the Efimov effect to provide a repulsive force.  When the Efimov resonance occurs, the three-body contribution dominates the energy density functional, leading to the stabilization of the system and the formation of the droplet. However, this multi-body system is shown to be very dissipative due to the three-body losses \cite{bulgac2002dilute}. 

In 1957, Lee, Huang and Yang \cite{lee1957eigenvalues} proposed the next-order energy correction for the ground state energy of weakly repelling Bose gas, i.e., the famous LHY correction, which can provide a repulsive force, $E/V=(gn^2/2)(1+128\sqrt{na^3}/15\sqrt{\pi}+\dots)$ with $a$ the scattering length and $n$ the particle number density. Usually, the ultra-cold atoms in the gas phase are trapped in the external potential and can easily expand into gas when the harmonic trap is switched off. Petrov was the first to propose \cite{petrov2015droplets} that diluted weakly interacting Bose mixtures can form droplets with self-binding properties. This attracted much attention from the cold atom community and inspired a huge wave of research in quantum droplet \cite{petrov2016lowdimensional,wang2020theory,ma2021borromean,li2018two,hu2020consistent,luo2021new,mistakidis2023few}.

Experimentally, the quantum droplets were observed for the first time in the dipolar system \cite{ferrier2016observation,schmitt2016self,guo2021new,malomed2020family,zheng2021quantum,chomaz2022dipolar} and the binary mixture system \cite{cabrera2018quantum,semeghini2018self,wilson2021quantum}. In both cases, the mechanism for the droplets is the same, i.e., the quantum droplet is stabilized by the precisely tuned attractive mean-field interactions and the repulsive quantum fluctuations. For the droplet formed by dipole interaction, the long- and short-range interactions compete at the mean-field level, and when trapped in a strong cigar-like potential, multiple droplets are arranged in a lattice pattern which breaks the rotational symmetry \cite{baillie2018droplet,norcia2021two}. In the binary mixture droplets, the residual mean field term as a combination of interspecies and intraspecies interactions can stabilize the LHY correction, and the magnetic Feshbach resonance is used to adjust the scattering length of atoms to achieve the required interaction strength for the self-binding mixture \cite{cabrera2018quantum,cheiney2018bright,semeghini2018self}. 

Collisions of quantum droplets have been studied in Bose–Bose mixtures with droplets prepared from separate condensates in a double-well potential \cite{ferioli2019collisions}. The collision outcome depends on the collision velocity, and the critical velocity that discriminates between different dynamical phases exhibits a different dependence on the atom number for small and large droplets. Using a time-dependent density functional theory, a numerical simulation of the collision of three-dimensional quantum droplets is carried out, which is consistent with the above experiment \cite{cikojevic2021dynamics} and the dynamics can be roughly divided into three situations: merging, separation, and evaporation. Compared with the quasi-elastic collision of Gaussian quantum droplets in free space, the slow-moving Gaussian quantum droplets in the shallow optical lattice potential have a tendency to merge after collision \cite{lao2021oscillatory}. The collision of one-dimensional (1D) droplets shows interesting breathing mode with the droplet size periodically oscillating \cite{astrakharchik2018dynamics}. The two-dimensional (2D) droplets are constructed with vorticity embedded into each component \cite{li2018two} and the merging of two zero-vorticity droplets into a single one with strongly oscillating eccentricity is briefly discussed. 

In this work, we study numerically the collisional behaviours in the 2D quantum droplets with equal interaction parameters and particle number density for each component. The quadrupole mode is found in the density width oscillation in the dynamics of the merged droplets. We manage to classify the dynamical phases into three catalogs and provide a criterion to determine the boundaries between them. 

The article is organized as follows. First, we introduce the characteristic units and rescale the dynamical equation into a dimensionless form in Sec.~\ref{Model}. In Sec. \ref{Collision dynamics of droplets}, the numerical scheme for the time-splitting method is introduced briefly. We discuss the quadrupole oscillation of the merged droplet and the damped oscillation of the separated droplet in Sec.~\ref{Merging phase}, and the oscillation period of the merged droplet is further analyzed. Finally, the phase diagram of the collisional dynamics is given in Sec.~\ref{Phase diagram} and we summarize the results in Sec.~\ref{conclusions}.

\section{Model}\label{Model}

We consider the 2D binary BEC with mutually symmetric components, assuming that the scattering length describing the intraspecies interaction is equal for each component, $a_{\uparrow\uparrow}=a_{\downarrow\downarrow}=a$, and the particle number density in the components is also equal, $n_{\uparrow}=n_{\downarrow}=n$. The corresponding energy density is \cite{petrov2016lowdimensional}
\begin{eqnarray}
	E_{2D}=\frac{8\pi \hbar^2n^2}{m\ln^2{(a_{\uparrow\downarrow}/a)}}[\ln(n/n_0)-1],  
\end{eqnarray}
with $n_0$ the equilibrium density of each component
\begin{eqnarray}
	n_0=\frac{e^{-2\gamma-3/2}}{2\pi}\frac{\ln(a_{\uparrow\downarrow}/a)}{aa_{\uparrow\downarrow}},
\end{eqnarray}
where $\gamma\approx0.5772$ is the Euler's constant. The stationary Gross-Pitaevskii equation is obtained by minimizing the energy functional with a non-uniform energy density $n(x,y)=|\psi(x,y)|^2$ with respect to independent variations of the wave function $\psi$ and its complex conjugate $\psi^*$ subject to the conservation of the total particle number $N =\int n(x,y) dxdy$ for a single component. To treat dynamical problems it is natural to use a time-dependent generalization of this Schr\"odinger equation, with the same non-linear interaction term
\begin{eqnarray}
	i\hbar\frac{\partial \psi}{\partial t}=\left[ - \frac{\hbar ^2}{2m}\nabla ^2 + \frac{8\pi\hbar^2}{m\ln^2(a_{\uparrow\downarrow}/a)}|\psi|^2\ln(\frac{|\psi|^2}{\sqrt{e}n_0}) \right]\psi, 
\end{eqnarray}
which is the basis for our discussion of the dynamics of the droplet. We define the characteristic units of length and time
\begin{eqnarray}
	{x_{\rm{0}}} &=& \sqrt {\frac{{\ln({a_{ \uparrow  \downarrow }}/a)a{a_{ \uparrow  \downarrow }}}}{{4{e^{ - 2\gamma  - 1}}}}},\\ 
	{t_0} &=& \frac{m{\ln({a_{ \uparrow  \downarrow }}/a)a{a_{ \uparrow  \downarrow }}}}{{4\hbar{e^{ - 2\gamma  - 1}}}}, 
\end{eqnarray}
which yields an energy unit 
\begin{eqnarray}
	{E_0} =\frac{\hbar^2}{m x_0^2}=\frac{\hbar}{t_0}= \frac{{4\hbar^2{e^{ - 2\gamma  - 1}}}}{m{\ln({a_{ \uparrow  \downarrow }}/a)a{a_{ \uparrow  \downarrow }}}}, 
\end{eqnarray}
a wave function normalization factor
\begin{eqnarray}
	{\psi _0} = \sqrt{\sqrt e {n_0}}= \sqrt {\frac{{{e^{ - 2\gamma  - 1}}\ln({a_{ \uparrow  \downarrow }}/a)}}{{2\pi a{a_{ \uparrow  \downarrow }}}}}, 
\end{eqnarray}
and a critical particle number 
\begin{eqnarray}
	{N_0} = \psi_0^2 x_0^2= \frac{\ln^2(a_{ \uparrow  \downarrow }/a)}{8\pi }. 
\end{eqnarray}
Thus, by rescaling the time, length, and wave function in these units $t= t't_0, x= x'x_0, \psi=\psi' \psi_0$, one obtains the dimensionless equation (with the primes omitted) that describes the dynamics of 2D quantum droplet 
\begin{eqnarray}
	i\frac{\partial  \psi }{\partial  t} = \left[ - \frac{ \nabla^2}{2} + \left| \psi  \right|^2 \ln(\left|  \psi  \right| ^2) \right] \psi. 
	\label{gpe2}
\end{eqnarray}
The spatial profile of the droplet $n(x, y)$ can be obtained by using the imaginary-time propagation method to solve Eq. (\ref{gpe2}) in the framework of the mean-field theory. This is governed by a single parameter, the dimensionless $N$, which is the number of particles divided by $N_0$. It has been shown \cite{li2018two} that the critical particle number $N_c$ separates two different physical regimes: for smaller $N$, the density profile is essentially nonuniform quasi-Gaussian as the quantum pressure is significant, while for larger $N$ a flat plateau is formed in the center of the droplet with the pattern similar to a puddle filled by the homogeneous liquid. The central density value of the droplet attains the maximum value $n_{\rm max} \simeq 0.6567$ in unit of $\psi_0^2$ at the border $N_c \simeq 48$ between the quasi-Gaussian and flat-top shapes, and gradually drops to the equilibrium value $n^{\rm TF}=1/\sqrt{e}$ estimated in the Thomas-Fermi approximation. It is worth to note that a similar puddle phase appears in the 1D droplet \cite{petrov2016lowdimensional,parisi2020quantum,mistakidis2021formation}, however, the central density value is monotonically increased to its equilibrium value $n_{\rm max} = 4/9$ in the unit of $\psi_0^2$ when $N$ goes infinity.

\section{Collision dynamics of droplets} \label{Collision dynamics of droplets}

\begin{figure*}
    \centering
    \includegraphics[width=14cm]{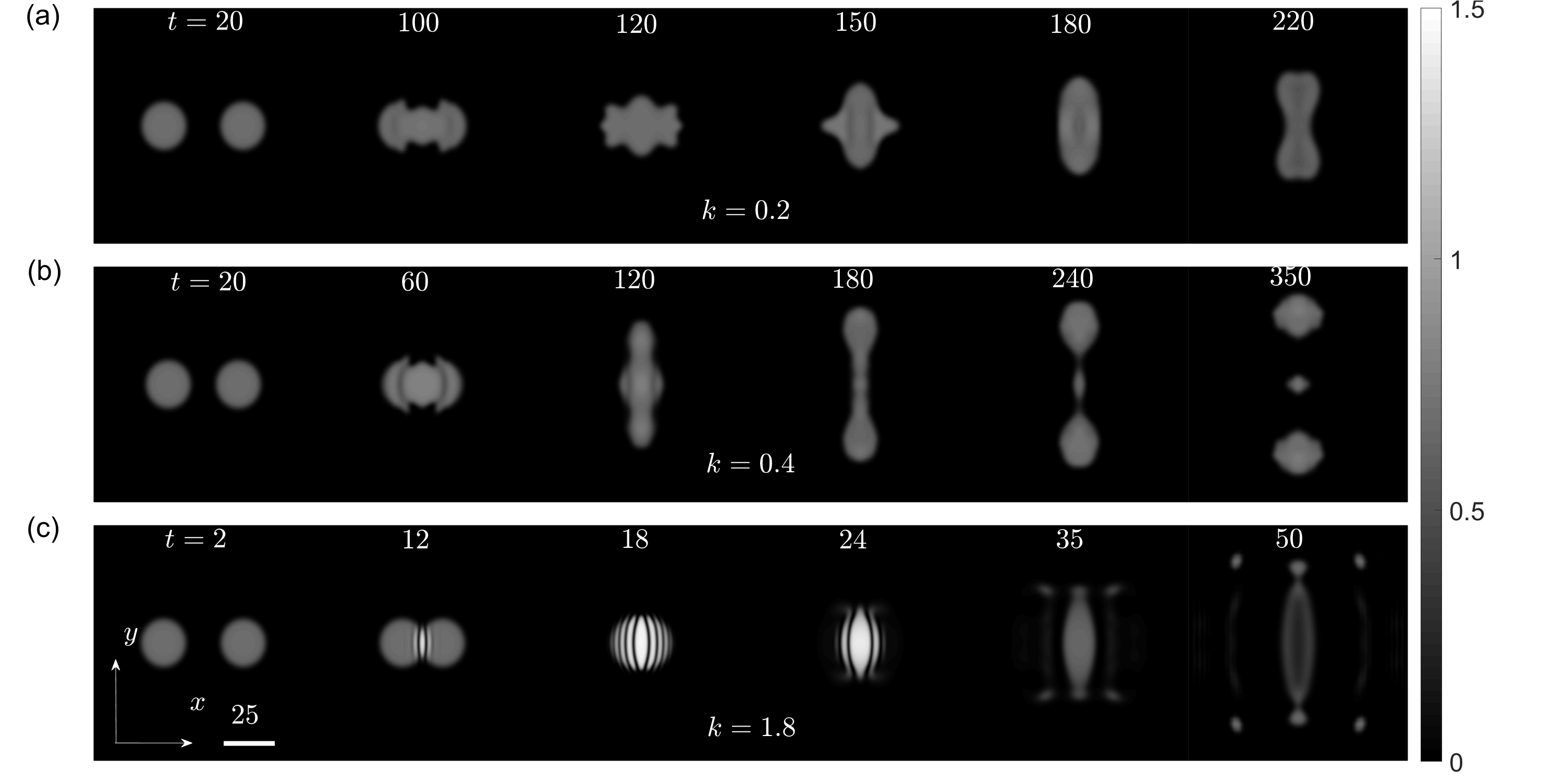}
    \caption{The density plots of the collisional dynamics of two droplets showing three different phases after the collision: (a) merging, (b) separation, and (c) evaporation, depending on the relative incident momentum $k = 0.2, 0.4$, and $1.8$, respectively, from top to bottom panels. Both droplets are initially normalized to a total particle number $N_1 = N_2 = 200$ and all three cases correspond to the in-phase collisions with $\phi=0$.}
    \label{Fig1}
\end{figure*}
The dynamics of quantum droplets exhibits the behavior similar to another type of
self-bound object, the so-called soliton \cite{baizakov2004multidimensional}, which can maintain its original shape without changing while traveling at a constant speed. Here we consider the pairwise collision between two droplets moving along the $x$ direction. The GPE (\ref{gpe2}) is usually solved by a time splitting method based on the fast Fourier transformation (FFT)~\cite{lehtovaara2007solution}: After a tiny time step $\Delta t$, the wavefunction of GPE (\ref{gpe2}) evolves as $\psi(x,y,\Delta t)=e^{-i\hat{H}\Delta t}\psi(x,y,0)$. Thus, the wavefunction at an arbitrary time $t=n\Delta t$ can be then written approximately as $\psi(x,y,t)=[e^{-i\hat{H}\Delta t}]^n\psi(x,y,0)$. For each time step, we use the Strang splitting $e^{-i\hat{H}\Delta t}=e^{-i(\hat{T}+\hat{V})\Delta t}\approx e^{-i\hat{T}\Delta t/2}e^{-i\hat{V}\Delta t}e^{-i\hat{T}\Delta t/2}$ to split the operator $e^{-i\hat{H}\Delta t}$ in terms of the kinetic part $\hat{T}$ and the potential $\hat{V}$. We then employ the FFT to deal with the operator $e^{-i\hat{T}\Delta t/2}$ in the momentum space while the diagonal potential $\hat{V}$ can evolve directly in the spatial space. We adopt the initial wave function in the superposition state of two droplets propagating in opposite directions
\begin{eqnarray}
	\psi(x,y,t=0)&=&\psi_1(x-a,y)e^{-ikx/2+i\phi}\notag\\&+&\psi_2(x+a,y)e^{+ikx/2},
\end{eqnarray}
where $\psi_{1,2}$ are the stationary shapes of quantum droplets with normalization particle number $N_{1,2}$, $\pm a$ are their initial positions in the $x$ direction, $k$ is the initial relative momentum of the colliding droplets, and $\phi$ is the relative phase. For simplicity, we consider the in-phase dynamics with $\phi=0$ of two droplets with equal number of particles $N_1=N_2$.

In Fig. \ref{Fig1} the collisional dynamics of two colliding droplets in 2D is presented for some typical values of relative momentum, for relatively large droplets initially normalized to a total particle number $N_1 = N_2 = 200$. In each panel, with increasing time from left to right, the density plots are shown for several moments of the collision process. Both droplets have the flat-top profile before the collision \cite{astrakharchik2018dynamics} due to their unique self-binding property, which can exist stably in the absence of an external potential until they meet each other. We see interference patterns appear when the two droplets meet and depending on the relative momentum, there exist three kinds of final states. For the collision with a small velocity $k = 0.2$, as shown in Fig. \ref{Fig1}(a), they merge into one droplet which is unevenly distributed and will still undergo a large amount of deformation as the liquid does. When the relative velocity is increased to $k = 0.4$, the droplets collide and interference occurs in one direction and then the atoms are separated into two droplets drifting away slowly in the vertical direction, leaving few atoms in the collisional center as shown in Fig. \ref{Fig1}(b). For even larger velocity $k = 1.8$ in Fig. \ref{Fig1}(c), we observe a strong interference pattern in the first stage of the collision and the droplets are then smashed into very small pieces without forming stable droplets, corresponding to the evaporation of liquid. The dynamics is calculated in a 2D space $L\times L$ with dimension $L=100x_0$ and the atoms can never reach the boundary in the evolution time in all panels of Fig. \ref{Fig1}. We can see that the density distribution is always symmetric about $x$ and $y$-axis during its evolution, which is guaranteed by the conservation of momentum.

\begin{figure}
    \centering
    \includegraphics[width=0.48\textwidth]{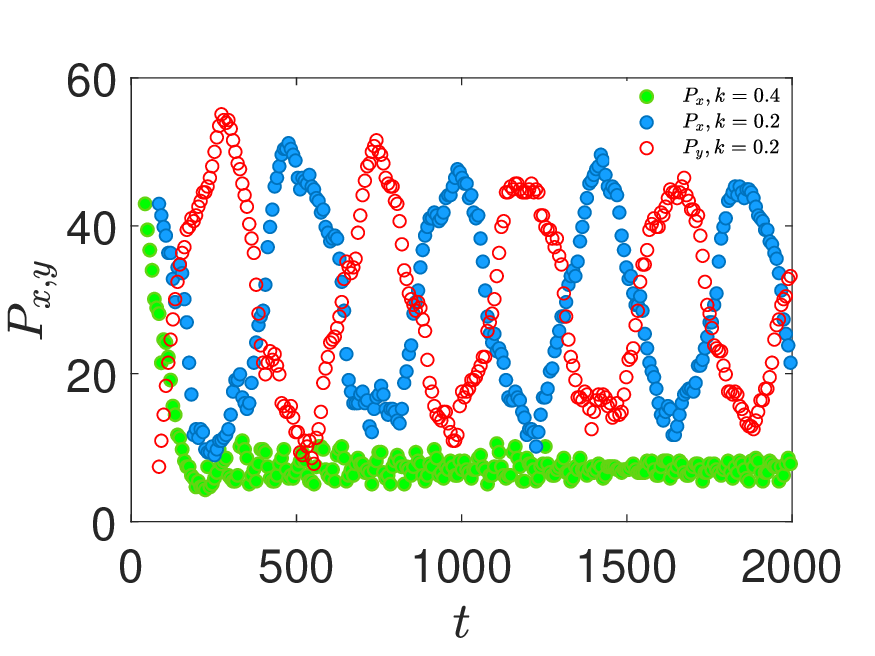}
    \caption{Typical quadrupole mode oscillation of the droplet in the merging phase and the damped oscillation in the separation phase. In the merging case ($k=0.2$) the droplet widths $P_x$ (blue solid circle) and $P_y$ (open red circle) oscillate periodically with time. In the separation case ($k=0.4$) the oscillation in $P_x$ (green solid circle) is quickly damped after the collision and the small fluctuation never exceeds half of the maximum value of droplet width $P_x$. Here $N_1=N_2=200$.}
    \label{Fig2}
\end{figure}

\section{Merging phase}\label{Merging phase}

It is interesting to further study the deformation of the merged droplet in the case of a slow collision. It is easy to see that the profile of the merged droplet experiences quadrupole mode oscillations in the $x$ and $y$ directions. To characterize the details we define the droplet width in the $x$ direction as the full width at half maximum (FWHM) of the probability density on the central line of $y=0$, i.e.,
\begin{eqnarray}
	P_x=x_{L}-x_{R} 
\end{eqnarray}
where $x_{L,R}$ denotes the left and right boundaries of the droplet. Similarly, the width in the $y$ direction is defined as the FWHM of the probability density on the central line of $x=0$, i.e.,
\begin{eqnarray}
	P_y=y_{T}-y_{D} 
\end{eqnarray}
where $y_{T,D}$ denotes the top and down boundaries of the droplet. We plot the shape oscillation in both directions in Fig. \ref{Fig2} for $k=0.2$ and $N_1=N_2=200$ and find that the merged droplet will never arrive at the boundary of our numerical simulation in the evolution and the quadrupole oscillation is restricted in a finite regime. The widths in both $x$ and $y$ directions, $P_x$ and $P_y$, oscillate periodically with time and alternately reach their maximum values as shown in Fig. \ref{Fig1}(a), i.e., $P_y$ is small when $P_x$ is large, and vice versa. The oscillation amplitudes of the merging droplets show a tendency to decay very slowly, as shown by the red and blue circles in Fig.~\ref{Fig2}. On the contrary, the oscillation in the separation phase is quickly damped after the collision and never exceeds half of the maximum value of droplet width $P_x$. The small fluctuation represents the shape oscillation of the remnant atoms in the origin area, while the two separated, bigger droplets drift away slowly along the orthogonal $y$-axis as shown in Fig. \ref{Fig1}(b).
\begin{figure}[tbp]
    \centering
    \includegraphics[width=0.48\textwidth]{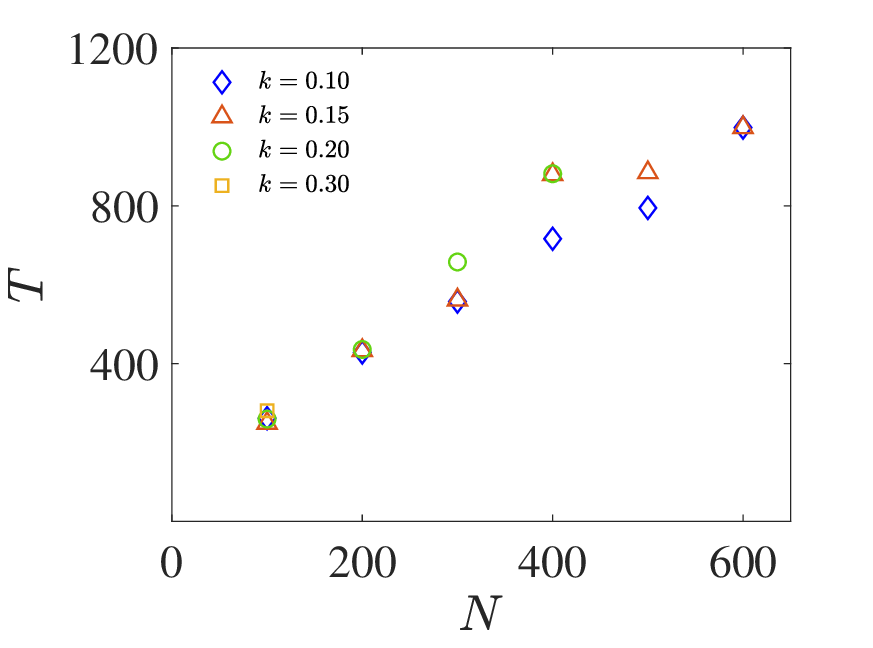}
    \caption{Period of quadrupole oscillation as a function of the particle number in the merging region for different incident momenta $k=0.1, 0.15, 0.2, 0.3$. For $k=0.3$, the colliding droplets will not merge for $N>150$. The calculation is done in a 2D space with a sufficiently large length $L=100x_0$ to assure the accuracy for large $N$. }
    \label{Fig3}
\end{figure}

The ultracold atom experiments allow one to witness how many-body effects emerge in the system as one gradually increases the particle number. Hence we further closely inspect the quadrupole oscillation period for various numbers of particles and different incident momenta in the merging region. We first let the colliding droplets merge into one and evolve for a long enough time. For each value of $k$ and $N$, the oscillation period is extracted from the quadrupole mode oscillation such as in Fig. \ref{Fig2} through the Fourier analysis. Specifically, the oscillation of $P_x$ for evolution time $t=4000$ is decomposed into its frequency components which is known as the frequency spectrum. The Fourier transform is represented as
\begin{eqnarray}
\tilde P_x(\omega) = \int_0^{+\infty} dt \left( P_x (t)- \bar P_x\right) e^{-i \omega t}
\end{eqnarray}
where $\tilde P_x(\omega)$ and $P_x (t)$ are the output and input spectra that are functions of frequency and time, respectively. The equilibrium value of the oscillation $\bar P_x$ averaged over the whole evolution time is subtracted to avoid the trivial peak at zero frequency. For a regular oscillation in our case, the frequency spectrum is distributed around a peak value $\omega_{\rm max}$. The period corresponding to the quadrupole oscillation is related to the frequency peak by $T =2\pi/\omega_{\rm max}$. We show the dependence of the quadrupole oscillation period as a function of the particle number in Fig. \ref{Fig3} for different incident momentum $k$. We scan the merging region for slow collision and make sure the colliding droplets merge into a bigger one in order to induce the quadrupole mode oscillation. Generally speaking, larger droplets have longer oscillation periods. For small droplets with $N<200$, the oscillation period seems to be independent of the incidental momentum $k$, as can be seen from the symbol overlap for $N=100$ and $200$ extracted from the oscillation of $P_x$. The period of the medium-size droplets with $N=300-500$ shows a tendency to increase with $k$, however, for even larger droplets ($N>600$) the collision is close to the separation phase and the oscillation period becomes ill-defined. The increase of oscillation period with incidental velocity is easily understood that the atoms in merged droplets may fly farther for a violent collision in the real time-of-flight process which surely takes longer time to finish a period before the self-binding interaction draws them back.

\begin{figure}[tbp]
    \centering
    \includegraphics[width=0.48\textwidth]{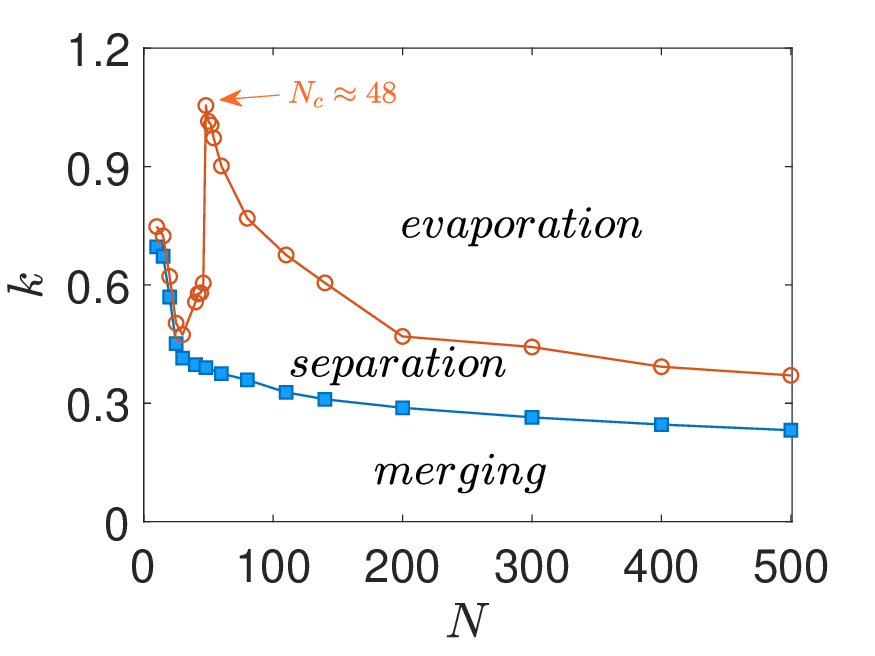}
    \caption{Phase diagram of the in-phase collisional dynamics of two droplets with equal particle number $N_1 = N_2=N$ and $\phi =0$. The boundaries between merging, separation, and evaporation phases are determined by the remnant particle number in the central area and the damped oscillation of the quadrupole mode, respectively. A stability peak for the existence of droplets emerges at the critical value $N_c \simeq 48$ for the quasi-Gaussian and flat-top shapes of the droplets.}
    \label{Fig4}
\end{figure}

\section{Phase diagram}\label{Phase diagram} 

We present in Fig. \ref{Fig4} the phase diagram of the collisional dynamics in terms of two parameters, i.e., the incidental momentum $k$ and the particle number $N$ in each droplet (for simplicity we again assume the collision occurs for two droplets with equal particle number $N_1=N_2=N$ and the relative phase before the collision is zero $\phi =0$). The behaviors of the collisions are classified into three situations: merging, separation, and evaporation, and the boundaries between them are determined by the damped oscillation of quadrupole mode in $P_x$, the remnant particle number in the central area, as well as the persistence of self-binding profile of the droplet after the collision. The most obvious difference between the separation and merging phases is the quadrupole oscillation of $P_{x,y}$ after the collision: in the merging case the droplet experiences a quadrupole oscillation in the area around the origin, while in the separation case the droplets drift away leaving behind a very small portion of atoms in the origin such that the droplet width (here we choose $P_x$) after the collision undergoes a very quick damping. For a fixed particle number, the critical momentum of the merging and separation regimes is determined by the oscillation amplitude of $P_x$ not exceeding a critical value, which is set as half of the difference between the maximum and minimum value of $P_x$ after the collision (see Fig. \ref{Fig2}).

On the other hand, the boundary between separation and evaporation phases is fixed by $R$, the ratio of the number of remnant particles in a selected central area $0.25L\times 0.25L$ in the $xy$-plane to the total number of particles. The remnant particle number is obtained by the integration in the chosen square area, namely $3L/8\le x,y \le 5L/8$, and $R=20\%$ serves as the criteria of evaporation-separation transition. One can see from the dynamics that the two separated droplets slowly move to the up and down borders leaving in the origin a small portion of atoms, see for instance Fig.~\ref{Fig1}(b). The atoms in the central area grow with increasing incidental momentum showing multiple droplets configuration, and finally the tiny pieces of the colliding droplets in the evaporation case are scattered all over the $xy$-plane, as shown in Fig.~\ref{Fig1}(c). When the ratio of atoms in the chosen square area becomes less than $20\%$, the droplet dynamics is classified into the evaporation phase. In practice, the central area integration is performed at the moment when $0.1\%$ of the total atoms have arrived at the square boundary of the system. The calculation is done in the $xy$ plane with a varying length $L$ which is taken as $8P_x$ for all the calculations. Thus, two phase boundaries may be determined accurately without the border reflection after the collision. We take the length of $L$ varing from $32.8x_0$ for $N=10$ to $253.1x_0$ for $N=500$ in Fig. \ref{Fig4}.

The phase diagram in Fig. \ref{Fig4} indicates that for slow collision the droplets prefer to merge into a bigger one with the quadrupole mode oscillation. When the incidental momentum is increased, the colliding droplets will separate into two droplets drifting along the vertical direction. For even larger momentum the collision will destroy the droplets into pieces scattering in the plane. The critical values of momentum of merging-separation (i.e., blue squares) decrease with the number of particles in the droplets and continue to approach a constant value. That is to say, for small droplet one needs larger incidental momentum to separate it into two or smashed into many pieces. In sharp contrast, the critical momentum of separation-evaporation (i.e., orange circles) exhibits instead a non-monotonic behaviour with respect to particle number. As $N$ decreases, the critical momentum increases monotonically first, reaches a maximum at $N_c \simeq 48$, and then reduces until almost joining the boundary between the merging and separation phases for a relatively small particle number. Similar to the case in one dimension~\cite{astrakharchik2018dynamics}, a stability peak is also found at $N_c \simeq 48$ for the collisions of the 2D droplets considered in this work. For even small droplets (i.e., $N=10\sim30$), the regime of the separation phase is significantly squeezed and the system enters directly from the merging phase into the evaporation phase with increasing $k$. It should be clarified that the boundary between separation and evaporation is fixed by the atoms in the central square area which is chosen artificially and the percentage of the remnant particles is also subject to vary. The boundaries and thus the phase diagram are not physically accurate and will change for alternative choice of these two parameters. Whether the colliding droplets merge or separate into multiple droplets depends on whether the surface tension is sufficient to offset the kinetic energy of the collision pair \cite{ashgriz1990coalescence,qian1997regimes,pan2005numerical,guo2021new} and the surface tension accounts for the droplet shape recovery which is independent of the droplet size in one dimension~\cite{astrakharchik2018dynamics}.

\section{Conclusions} \label{conclusions}

We have studied the collisional dynamics of two symmetric quantum droplets in two dimensions by solving the extended nonlinear Schr\"odinger equation with the inclusion of the LHY term. The dynamics shows three different phases after the collision, i.e. merging, separation, and evaporation. The phase diagram is obtained in the parameter space of incidental momentum and the particle number. In the merging phase, the droplet shape exhibits typical quadrupole mode oscillation, in the separation phase the oscillation is damped very quickly, and in the evaporation phase the remnant number of particles in the central square area amounts to a considerable ratio. This allows us to distinguish them easily. The dynamics for different particle numbers in the droplets, with non-zero relative phase, and for non-head-on collision may exhibit more complicated patterns that deserve further studies.  The difference between smaller and larger droplets corresponds to the crossover from a compressible state to an incompressible state where the dynamics is dominated by the droplet binding energy and the surface tension respectively. 

\begin{acknowledgments}
The authors are grateful to Dr. Li Chen for illuminating discussions on FFT. This work is supported by the National Natural Science Foundation of China (Grant No. 12074340) and the Science Foundation of Zhejiang Sci-Tech University (ZSTU) under Grants No. 20062098-Y and No. 21062339-Y.
\end{acknowledgments}

\bibliography{dropletscollision2d}

\end{document}